\documentclass[%
 reprint,amsmath,amssymb,
 aps,]{revtex4-1}

\usepackage{graphicx}% Include figure files
\usepackage{dcolumn}% Align table columns on decimal point
\usepackage{bm}% bold math

\begin{document}
\title{Generalized effective-medium theory for metamaterials}
\author{Brian A. Slovick, Zhi Gang Yu, and Srini Krishnamurthy}
\address{Applied Optics Laboratory, SRI International, 333 Ravenswood Avenue,
Menlo Park, California 94025, USA}
\date{\today}
\begin{abstract}
We present an effective-medium model for calculating the frequency-dependent effective permittivity $\epsilon(\omega)$ and permeability $\mu(\omega)$ of metamaterial composites containing spherical particles with arbitrary permittivity and permeability. The model is derived from the zero-scattering condition within the dipole approximation, but does not invoke any additional long-wavelength approximations. As a result, it captures the effects of spatial dispersion and predicts a finite effective refractive index and antiresonances in $\epsilon(\omega)$ and $\mu(\omega)$, in agreement with numerical finite-element calculations.
\end{abstract}
 \maketitle

\section{\label{sec:level1}Introduction\protect\\}
Metamaterials (MMs) possessing unusual values of the electric permittivity $\epsilon(\omega)$ and magnetic permeability $\mu(\omega)$ exhibit fascinating responses to electromagnetic waves, promising new physics and many novel applications \cite{Soukoulis2011,Zheludev2012}. Reliable and efficient modeling tools play an indispensable role in understanding the physics and advancing the field of MMs. One approach is to discretize the system using the finite-element method \cite{Cui2009} and employ $S$-parameter retrieval \cite{Smith2002}. While nearly exact, this method is limited to ordered arrays and is time-consuming and often ambiguous in extracting the effective  $\epsilon(\omega)$ and $\mu(\omega)$ of a composite \cite{Chen2004,Smith2005}. Another approach is to develop analytical  effective medium (EM) models \cite{Lewin1947,Stroud1978,Niklasson1981,Bohren1986,Doyle1989,Stroud1998,Wheeler2005,Wu2006,Kuester2011}. While these models are simple and efficient, their validity and accuracy are limited by the long-wavelength approximation, requiring the length of the MM unit cell to be much smaller than the wavelength in the effective medium \cite{Kuester2011}. Moreover, the existing EM models do not adequately account for the spatial dispersion in inhomogeneous MMs, of which one consequence is the frequently observed antiresonance \cite{Koschny2003,Koschny2004,Starr2004}, whereby a resonance in $\epsilon$ (or $\mu$) is accompanied by an inverted resonance in $\mu$ (or $\epsilon$) at the same frequency. Such antiresonances are absent in existing EM theories where the frequency-dependent $\epsilon(\omega)$ and $\mu(\omega)$ are decoupled. Another serious deficiency in existing EM models is the lack of self-consistency. For example, the EM models developed by Lewin \cite{Lewin1947} and Wu et al. \cite{Wu2006} predict an infinitely large effective refractive index $n$ ($=\sqrt{\epsilon\mu}$) near the resonances, leading to a wavelength of zero in the effective medium, contradicting the long-wavelength approximation used to derive the models.

In this article, we derive a generalized effective medium (GEM) model, along the lines of Lewin and Wu, for metamaterial composites containing spherical particles of arbitrary permittivity and permeability. By appropriate change of variables, we obtain two decoupled equations, one closed-form analytical expression for the impedance $z$ ($=\sqrt{\mu/\epsilon}$) and a nonlinear equation for $n$, requiring a straightforward numerical solution. As a result of this decoupling, the uniqueness and stability of the solution are guaranteed. Moreover, the effective $\epsilon$ and $\mu$ can be calculated without invoking the long-wavelength approximation used by Lewin and Wu. By comparison to finite-element numerical calculations, we further show that the GEM model is valid closer to the Mie resonances and over a much broader range of the constitutive permittivities compared to existing EM models. The GEM model correctly predicts a finite effective refractive index at the Mie resonances and antiresonances in the effective permittivity and permeability. We further demonstrate, without explicitly assuming periodicity, that the antiresonances in the GEM model originate from complementary resonances in the background region of the composite.

\section{\label{sec:level1}Approach\protect\\}
\begin{figure}
\includegraphics[width=45mm]{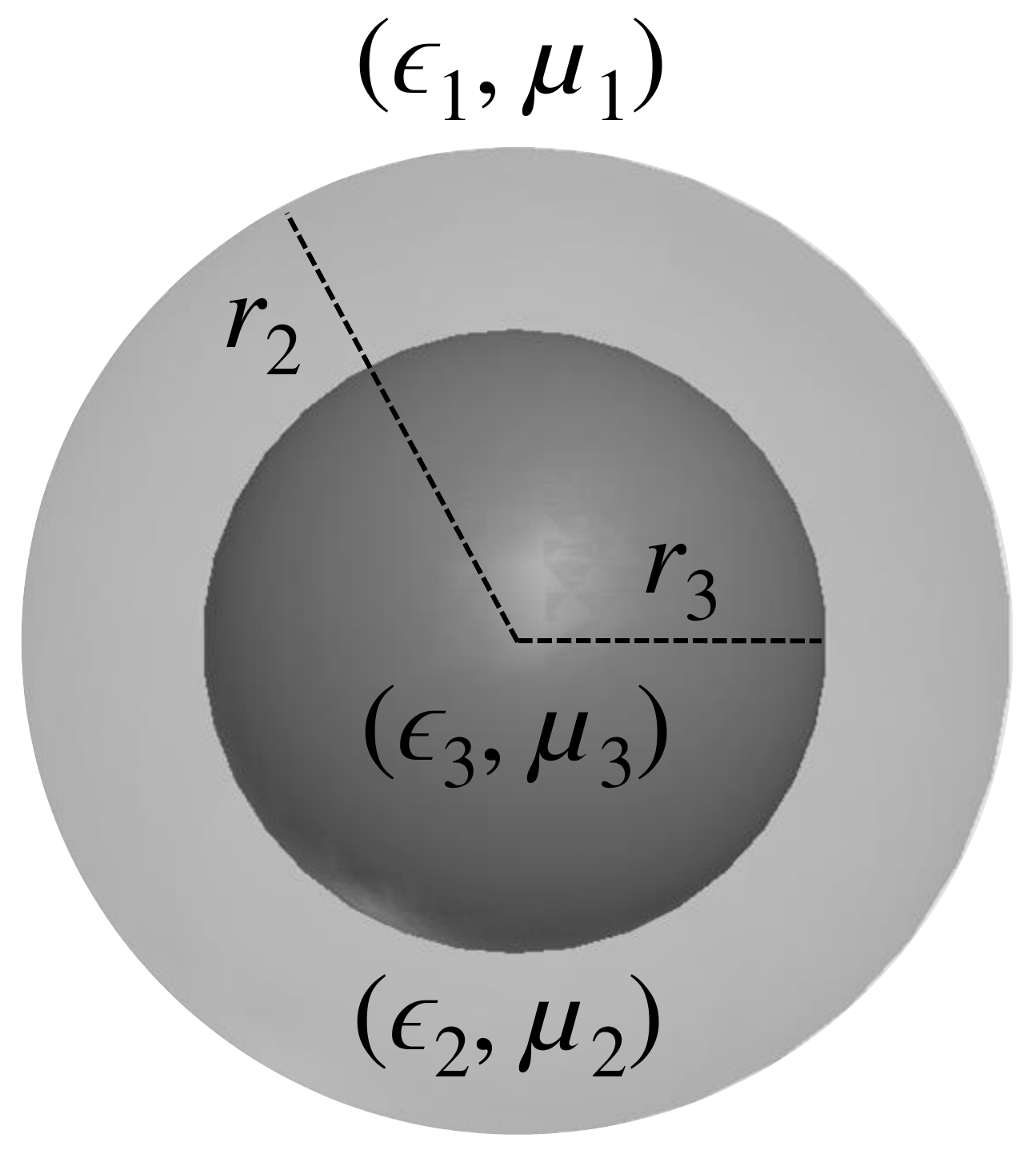}
\caption{\label{fig:epsart} Definition of terms for the GEM model. The parameters of the effective medium $\epsilon_1$ and $\mu_1$ are determined by the condition that electromagnetic plane waves incident from the effective medium on the core-shell structure do not scatter.}
\end{figure}

Similar to previous models \cite{Stroud1998,Niklasson1981,Wu2006}, the GEM model consists of magnetodielectric spheres of radius $r_3$ with arbitrary relative permittivity $\epsilon_3$ and relative permeability $\mu_3$, embedded in a background medium with $\epsilon_2$ and $\mu_2$ with a volume fraction $f$. The unit cell of the composite is a core-shell structure, as shown in Fig. 1, with outer radius $r_2$, determined from the condition that the ratio of the volume of the core to that of the shell is $f$ [$=(r_3/r_2)^3$]. The core-shell structure is embedded in an effective medium with $\epsilon_1$ and $\mu_1$ values that are determined by the condition that electromagnetic plane waves incident from the medium on the core-shell structure do not scatter. Retaining only the dipole terms, the scattering cross section of the core-shell structure is \cite{Bohren1983}
\begin{equation}
\sigma_{sca}=\frac{6 \pi}{k_1^2}(|a_1|^2+|b_1|^2),
\end{equation}
where $a_1$ and $b_1$ are the electric- and magnetic-dipole Mie-scattering coefficients of the core-shell \cite{Bohren1983,Kuester2011},
$$
a_1=\frac{\sqrt{\epsilon_2/\epsilon_1}\psi'_1(k_1r_2)G_\epsilon(k_2r_2)-\sqrt{\mu_2/\mu_1}\psi_1(k_1r_2)G'_\epsilon(k_2r_2)}{\sqrt{\epsilon_2/\epsilon_1}\xi'_1(k_1r_2)G_\epsilon(k_2r_2)-\sqrt{\mu_2/\mu_1}\xi_1(k_1r_2)G'_\epsilon(k_2r_2)}
$$
and
$$
b_1=\frac{\sqrt{\mu_2/\mu_1}\psi'_1(k_1r_2)G_\mu(k_2r_2)-\sqrt{\epsilon_2/\epsilon_1}\psi_1(k_1r_2)G'_\mu(k_2r_2)}{\sqrt{\mu_2/\mu_1}\xi'_1(k_1r_2)G_\mu(k_2r_2)-\sqrt{\epsilon_2/\epsilon_1}\xi_1(k_1r_2)G'_\mu(k_2r_2)},
$$
where $k_i=\omega/c\sqrt{\epsilon_i \mu_i}$ ($i=1,2,3$), $\psi_1(x)=zj_1(x)$ and $\chi_1(x)=-xy_1(x)$ are the Riccati-Bessel functions and $\xi_1(x)=\psi_1(x)+j\chi_1(x)$, with  $j_1(x)$ and $y_1(x)$ being the spherical Bessel functions and
$$
G_{\epsilon}(x)=\psi_1(x)-A_1\chi_1(x),~G_{\mu}(x)=\psi_1(x)-B_1\chi_1(x),
$$
where 
$$
A_1=\frac{\epsilon_3 F(k_3 r_3) k_2 r_3 \psi'_1(k_2 r_3)-2\epsilon_2 \psi_1(k_2 r_3)}
{\epsilon_3 F(k_3 r_3) k_2 r_3 \chi'_1(k_2 r_3)-2\epsilon_2 \chi_1(k_2 r_3)},
$$
 $B_1 = A_1 \{\epsilon_i\to \mu_i\}$  \cite{Bohren1983,Kuester2011}, and 
$$
F(x)=\frac{2(\sin{x}-x\cos{x})}{x\cos{x}+(x^2-1)\sin{x}}.
$$
Changing variables to $z_i=\sqrt{\mu_i / \epsilon_i}$ and setting $a_1$ and $b_1$ individually equal to zero [to obtain $\sigma_{sca}=0$], we obtain two conditions:
\begin{equation}
\frac{\psi_1(k_1r_2)}{\psi'_1(k_1r_2)}\equiv \frac{1}{2}k_1 r_2 F(k_1 r_2)=\left[\frac{G_{\epsilon}(k_2 r_2)G_{\mu}(k_2 r_2)}{G'_{\epsilon}(k_2 r_2)G'_{\mu}(k_2 r_2)}\right]^{1/2}
\end{equation}
and
\begin{equation}
z_1=z_2\left[\frac{G'_{\epsilon}(k_2 r_2)G_{\mu}(k_2 r_2)}{G_{\epsilon}(k_2 r_2)G'_{\mu}(k_2 r_2)}\right]^{1/2}.
\end{equation}
Noting that the right-hand sides (RHS) of Eqs. (2) and (3) are known for a given frequency, Eq. (2) can be solved numerically to obtain $k_1=\omega/c\sqrt{\epsilon_1 \mu_1}$. Together with $z_1=\sqrt{\mu_1/\epsilon_1}$, obtained from Eq. (3), the effective $\epsilon_1$ and $\mu_1$ are calculated from the expressions
\begin{equation}
\epsilon_1=n_1/z_1\quad \text{and} \quad\mu_1=n_1z_1.
\end{equation}
All effective parameters in the GEM model are obtained from Eqs. (2)-(4).

The physical significance of the function $F(k_1r_2)$, because it contains periodic functions of the phase shift, is related to diffraction (i.e., spatial dispersion). It is important to note that at the Mie resonances, where the RHS of Eq. (2) diverges, the GEM model predicts a finite $k_1$ (and therefore $n_1$) because $F(k_1r_2)$ has poles near $k_1r_2=2.7$ and 6.1. 

We now show how Eqs. (2) and (3) reduce to the expressions derived previously by Lewin and Wu. When the wavelength in the effective medium is much larger than the shell diameter (that is, $k_1r_2 \ll 1$), $F(k_1r_2)\simeq1$ and Eq. (2) reduces to
\begin{equation}
\frac{1}{2}k_1r_2=\left[\frac{G_{\epsilon}(k_2 r_2)G_{\mu}(k_2 r_2)}{G'_{\epsilon}(k_2 r_2)G'_{\mu}(k_2 r_2)}\right]^{1/2},
\end{equation}
while Eq. (3) is unchanged. By substituting Eq. (3) and (5) into Eq. (4), we obtain
\begin{equation}
\epsilon_1=\frac{2\epsilon_2}{k_2r_2}\frac{G_{\epsilon}(k_2 r_2)}{G'_{\epsilon}(k_2 r_2)}\quad \text{and} \quad\mu_1=\frac{2\mu_2}{k_2r_2}\frac{G_{\mu}(k_2 r_2)}{G'_{\mu}(k_2 r_2)},
\end{equation}
which are equivalent to the expressions derived by Wu et al. \cite{Wu2006}. Note that at the Mie resonances, where the RHS of Eq. (5) diverges, the left-hand side remains finite unless $n_1$ is made to diverge, resulting in an unphysical value. This divergence occurs because diffraction is absent [i.e., $F(x)=1$] in the long-wavelength limit.

If the additional approximation is made that the wavelength in the shell region is large compared to the shell diameter (that is, $k_2r_2 \ll 1$ and thus also $k_2r_3 \ll1$) while $k_3r_3$ remains arbitrary, the Riccati-Bessel functions can be replaced by their small-argument approximations \cite{Kuester2011} 
$$
\psi_1(x)\simeq\frac{x^2}{3};\quad \psi'_1(x)\simeq\frac{2x}{3};\quad \chi_1(x)\simeq\frac{1}{x};\quad \chi'_1(x)\simeq-\frac{1}{x^2}
$$
to obtain
\begin{equation}
\frac{G_{\epsilon}(k_2r_2)}{G'_{\epsilon}(k_2r_2)}=\frac{k_2r_2}{2}\frac{1+2f\frac{\epsilon_3F(k_3r_3)-\epsilon_2}{\epsilon_3F(k_3r_3)+2\epsilon_2}}{1-f\frac{\epsilon_3F(k_3r_3)-\epsilon_2}{\epsilon_3F(k_3r_3)+2\epsilon_2}},
\end{equation}
and a similar expression for $G_{\mu}(k_2r_2)/G'_{\mu}(k_2r_2)$. Substituting Eq. (7) into Eq. (6), we obtain
\begin{equation}
\epsilon_1=\epsilon_2\frac{1+2f\frac{\epsilon_3F(k_3r_3)-\epsilon_2}{\epsilon_3F(k_3r_3)+2\epsilon_2}}{1-f\frac{\epsilon_3F(k_3r_3)-\epsilon_2}{\epsilon_3F(k_3r_3)+2\epsilon_2}},
\end{equation}
and
\begin{equation}
\mu_1=\mu_2\frac{1+2f\frac{\mu_3F(k_3r_3)-\mu_2}{\mu_3F(k_3r_3)+2\mu_2}}{1-f\frac{\mu_3F(k_3r_3)-\mu_2}{\mu_3F(k_3r_3)+2\mu_2}},
\end{equation}
which are equivalent to the expressions obtained originally by Lewin \cite{Lewin1947}.

By retaining only the dipole terms, the GEM model, like the Lewin and Wu models, implicitly assumes long wavelengths. Since the quadrupole terms are proportional to the product of the dipole terms and $(k_1r_2)^2$ \cite{Kuester2011,Bohren1983}, by retaining only the dipole terms, all three models implicitly assume that $(k_1r_2)^2 \ll 1$. However, only the Lewin and Wu models make the additional assumption that $k_1r_2 \ll 1$. Therefore, the GEM model is expected to be valid to larger values of $k_1r_2$, i.e., to higher frequencies and closer to the Mie resonances where $n_1$ is large.

\section{\label{sec:level1}Validation\protect\\}

To assess the accuracy and range of validity of the GEM, Wu, and Lewin models, the results are compared to numerical finite-element calculations from the commercial code HFSS (Ansys), in which the effective parameters are calculated from the transmission and reflection coefficients for a single layer of the MM \cite{Smith2002,Smith2005}, with the particles arranged in a square lattice.

\begin{figure}
\includegraphics[width=88mm]{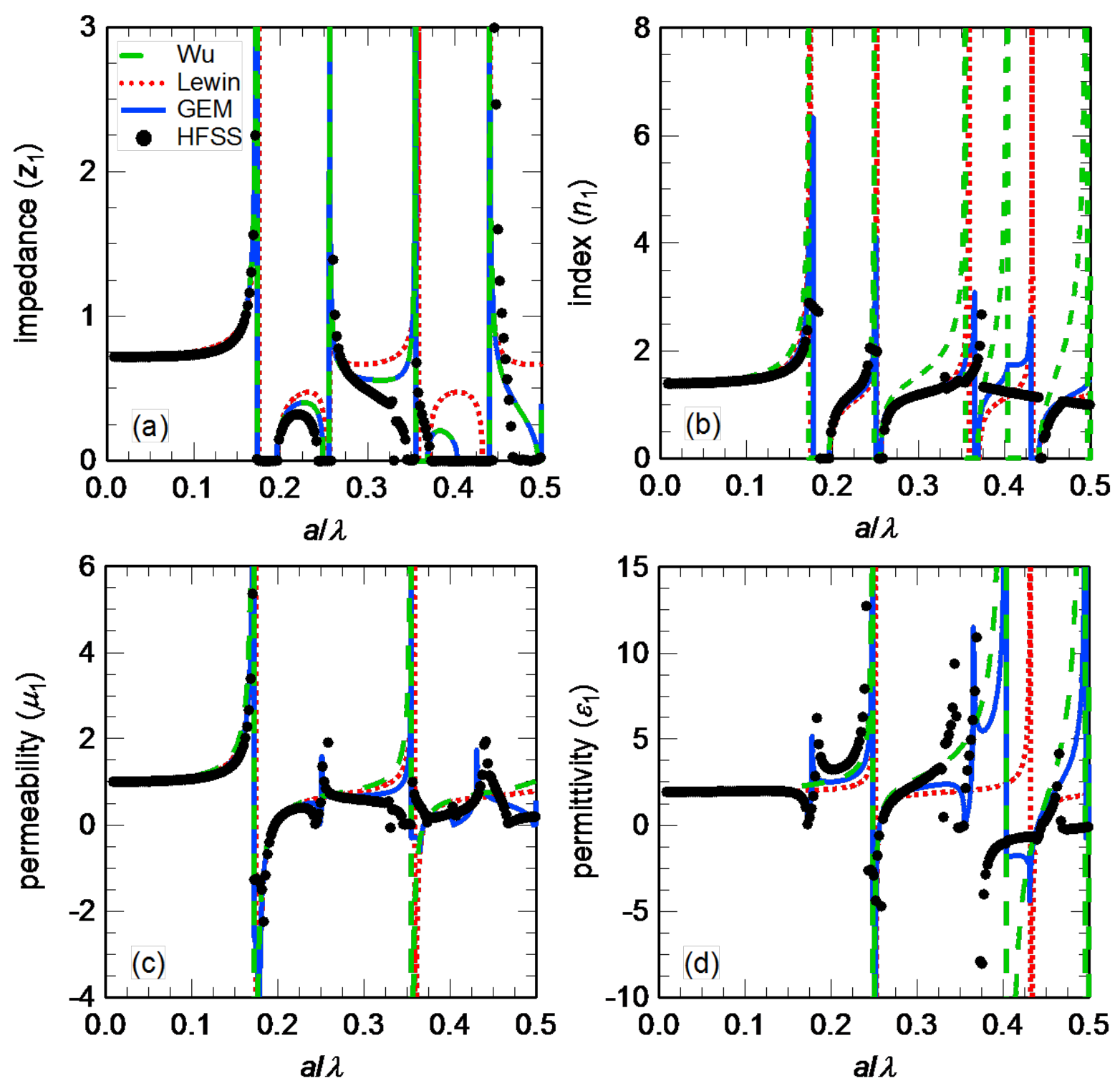}
\caption{\label{fig:epsart} Comparison of calculated real values of the effective (a) impedance, (b) index, (c) permeability, and (d) permittivity of the metamaterial composite of dielectric spheres with $\epsilon_3=50$ and $\mu_3=1$, embedded in vacuum with a volume fraction of $f=0.25$.}
\end{figure}

For illustration, first we consider dielectric spheres with large permittivity ($\epsilon_3=50$ and $\mu_3=1$) embedded in vacuum ($\epsilon_2=\mu_2=1$). The calculated impedance $z_1$ index $n_1$, $\epsilon_1$, and $\mu_1$ for this case are shown in Fig. 2 as a function of $a/\lambda$, where $a$ is the unit-cell length of the cubic lattice for the chosen $f$ and $r_3$. Owing to the large sphere permittivity, the Mie resonances occur at low frequencies, where the long-wavelength approximation used by Lewin and Wu is valid \cite{Kuester2011}.  As a result, at low frequencies all three models (as marked) produce nearly identical results and compare well with the finite-element calculations (symbols). The equation for $z_1$ is the same in both the GEM and Wu models [Fig. 2(a)], and differs from the Lewin model only for $a/\lambda>0.3$ (or $ka>1.9$), where all three models become inaccurate. Since the GEM model obtains $n_1$ directly from Eq. (2), a finite value for $n_1$ is predicted [Fig. 2(b)] for all frequencies. The calculated values, even at the resonances, are in good agreement with the HFSS values, whereas the other models predict either very large or infinite values, as described earlier. Since both GEM and HFSS obtain finite values for $n_1$, the increase in $\mu_1$ near the resonance at $a/\lambda\approx0.18$ is accompanied by a decrease, or antiresonance, in $\epsilon_1$ at the same frequency. Similarly, the resonance in $\epsilon_1$ at $a/\lambda\approx0.25$ generates an antiresonance in $\mu_1$ at the same frequency. This antiresonance feature is absent in the other models because they allow for infinite values of $n_1$ near the resonances. As a result, only the GEM model accurately predicts the effective $\epsilon_1$ and $\mu_1$ through the first two Mie resonances (i.e., for $a/\lambda<0.3$).

\begin{figure}
\includegraphics[width=85mm]{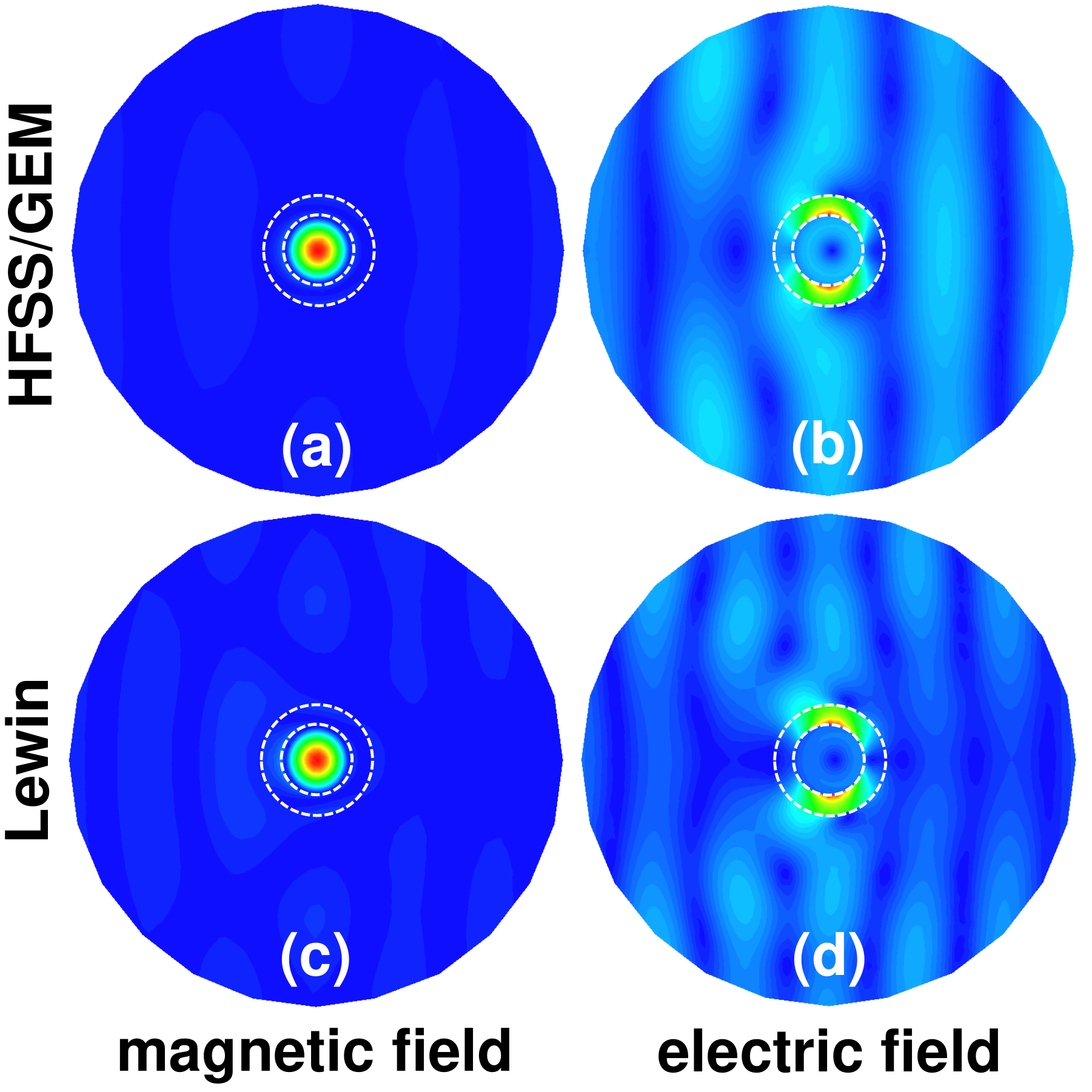}
\caption{\label{fig:epsart} Calculated magnetic and electric field around the core-shell described in Fig. 2, embedded in a medium with effective parameters $\epsilon_1$ and $\mu_1$ calculated in HFSS (a)-(b) (equivalent to GEM in this case) and the Lewin model (c)-(d), at a frequency just below the first resonance at $a/\lambda\approx0.17$, for a plane wave incident from the left.}
\end{figure}

Next we demonstrate that the effective $\epsilon_1$ and $\mu_1$ obtained in HFSS and the GEM model accurately represent the effective medium, and that both electric and magnetic resonances are simultaneously present at the same frequency.  We place a single core-shell structure in an infinite medium with effective parameters $\epsilon_1$ and $\mu_1$ calculated in HFSS (equivalent to the GEM model in this case) at a frequency $a/\lambda\approx0.17$, just below the first resonance in $\mu_1$ and antiresonance in $\epsilon_1$ (see Fig. 2). Using HFSS, we obtained the magnitude of the time-dependent electric and magnetic field in and around the core-shell structure (Fig. 3).  As seen in Fig. 3(a), the magnetic field clearly resembles a magnetic-dipole mode \cite{Shi2012}, with a single maximum at the center of the core. The electric field at the same frequency is nonuniform and mostly concentrated in the shell region [Fig. 3(b)], indicating that the magnetic resonance in the core is accompanied by an electric resonance in the shell. Since the HFSS calculation of the effective $\epsilon_1$ for this composite shows an antiresonance at this frequency, we conclude that the origin of the antiresonance in $\epsilon_1$ is an electric resonance in the background region of the composite. Figure 3(b) further shows that for an electromagnetic plane wave incident from the left, no shadow is cast in the transmitted (right) side of the core-shell, demonstrating that the HFSS (and GEM) values of $\epsilon_1$ and $\mu_1$ accurately represent the effective medium. We also carried out similar calculations with the effective parameters obtained from the Lewin model. In this case, as shown in Fig. 3(d), a shadow is clearly cast on the transmitted side of the core-shell structure, indicating that the effective parameters obtained from the Lewin model do not satisfy the zero-scattering condition, and thus are inaccurate compared to HFSS and GEM at this frequency.

From the GEM model, it is clear that the zero-scattering condition, treated consistently with resonances in the core and shell, leads to antiresonances, allowing the imaginary part of either $\epsilon$ or $\mu$ to be negative. To verify that the energy dissipation in the GEM model is positive, which is required to be consistent with thermodynamics \cite{Landau1984,Koschny2003}, we used the expression for the energy dissipation in a homogeneous medium \cite{Landau1984},
\begin{equation}
Q=\frac{\omega}{8\pi}\left[\text{Im}\epsilon(\omega)|E(\omega)|^2+\text{Im}\mu(\omega)|H(\omega)|^2\right], \nonumber
\end{equation}
and the condition for a plane wave $E(\omega)/H(\omega)=z(\omega)$ to calculate the energy dissipation for several cases with material loss (e.g., $\epsilon_3=50+j0.01$). In the cases considered, we found that $Q$ is always $\ge0$, even at the antiresonances. Because the imaginary parts of $\epsilon$ and $\mu$ are traditionally required to be positive, it has been suggested that antiresonances are artifacts resulting from the treatment of a spatially inhomogeneous medium by an effective homogeneous medium model \cite{Koschny2004,Efros2004,Alu2011a,Alu2011b}.

To further explore the range of validity of the GEM model, we now consider dielectric spheres with small permittivity ($\epsilon_3=12$ and $\mu_3=1$, corresponding to silicon in the short-wave infrared) embedded in vacuum. The calculated values of $z_1$, $n_1$, $\epsilon_1$, and $\mu_1$ are shown in Fig. 4, as a function of $a/\lambda$. In this case, the Mie resonances occur at higher frequencies ($a/\lambda\approx0.3$), where the long-wavelength approximations used by Lewin and Wu are suspect. In addition, the higher-order multipole terms ($a_m$ and $b_m$ with $m>1$), neglected in all three models, become significant for spheres with small permittivity, and may even overlap with the lower-energy dipole resonances \cite{Garcia-Etxarri2011,Shi2012}. Therefore, in this case all three models have limited agreement with HFSS. The $z_1$ values obtained by the GEM and Wu models are equivalent, agreeing well with HFSS up to $a/\lambda\approx0.3$ [Fig. 4(a)], and both models accurately predict the location of the resonance in $z_1$. But only the GEM model predicts a finite $n_1$ similar to HFSS [Fig. 4(b)]. As a result, the effective $\epsilon_1$ and $\mu_1$ calculated with the GEM model are closer to the HFSS values than the other models, predicting both the correct location and approximate magnitude of the first two Mie resonances.

\begin{figure}
\includegraphics[width=87mm]{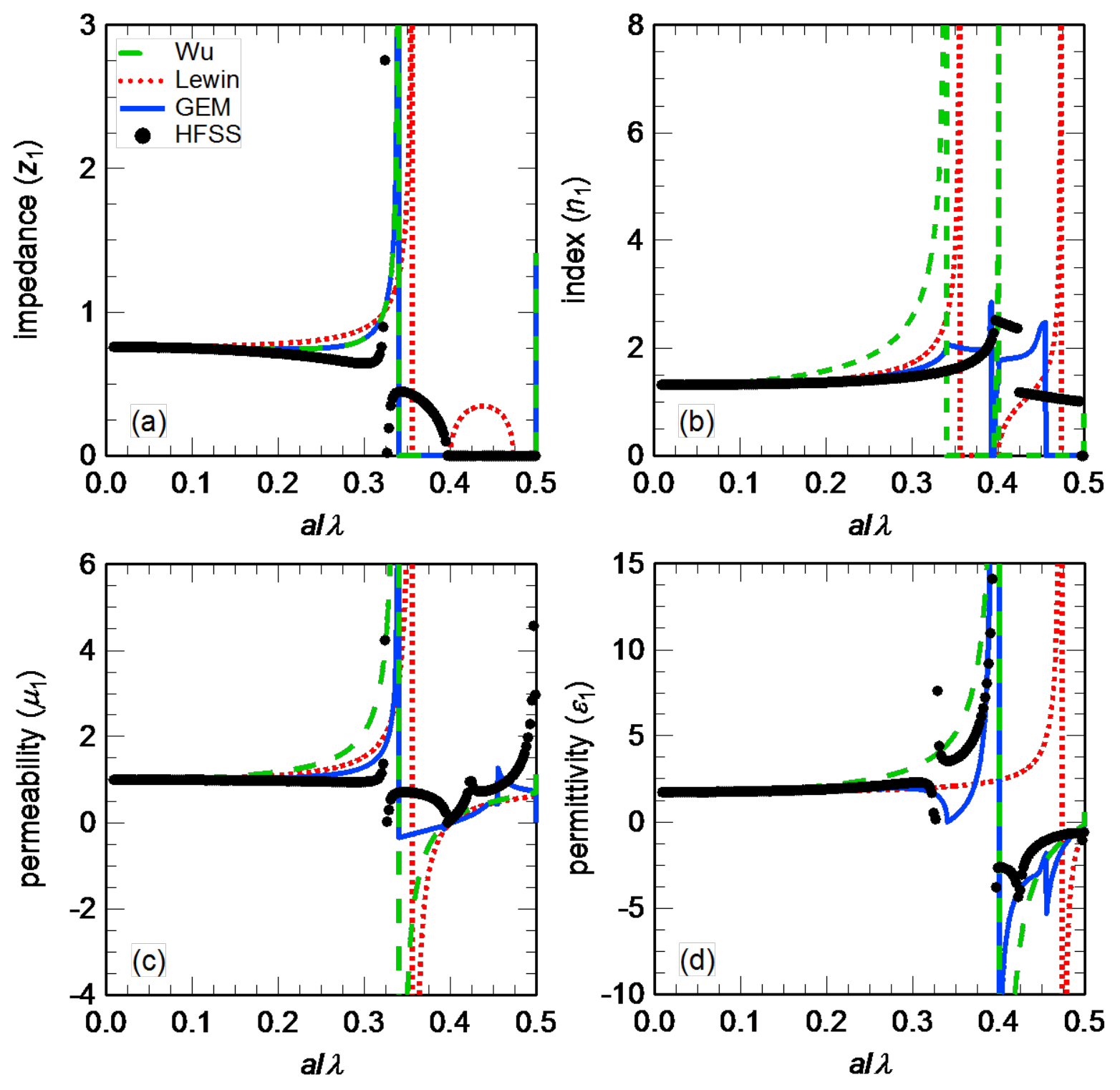}
\caption{\label{fig:epsart} Comparison of calculated real values of the effective (a) impedance, (b) index, (c) permeability, and (d) permittivity of the metamaterial composite of dielectric spheres with $\epsilon_3=12$ and $\mu_3=1$, embedded in vacuum with a volume fraction of $f=0.25$.}
\end{figure}

\begin{figure}
\includegraphics[width=87mm]{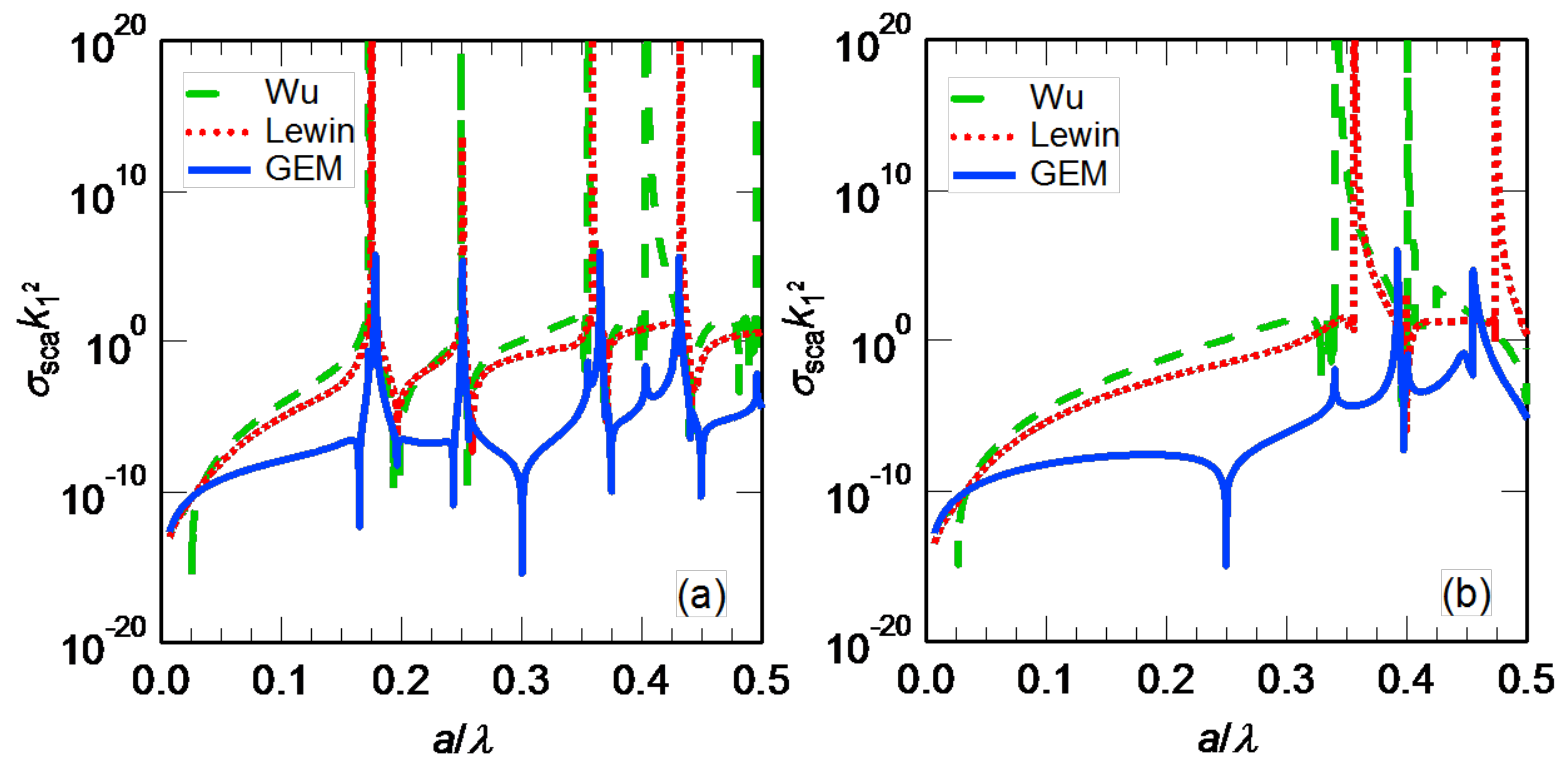}
\caption{\label{fig:epsart} Comparison of the normalized scattering cross sections, calculated from Eq. (1), for core-shells with $\epsilon_2=\mu_2=1$ and $f=0.25$ and (a) $\epsilon_3=50$ and $\mu_3=1$, and (b) $\epsilon_3=12$ and $\mu_3=1$, embedded in a medium with effective $\epsilon_1$ and $\mu_1$ values obtained from the three EM models.}
\end{figure}

To further illustrate the extended range of validity of the GEM model, using Eq. (1) we calculated the normalized scattering cross sections ($=\sigma_{sca} k_1^2$) for core-shells embedded in a medium with effective $\epsilon_1$ and $\mu_1$ values obtained from the Lewin, Wu, and GEM models (Fig. 5). Since, by definition, the scattering cross section of core-shells embedded in their effective medium is zero, the model with the smallest scattering cross section is the one that most accurately describes the effective medium. Assuming that a normalized cross section less than $10^{-3}$ accurately describes the effective medium, for spheres with large permittivity, corresponding to the case in Fig. 2, all three effective-medium models generally perform well [Fig. 5(a)], except very close to the Mie resonances, where $n_1$ is large and the wavelength in the effective medium is small. At virtually all frequencies, the GEM model has the smallest cross section, followed by the Lewin and Wu models. For spheres with smaller permittivity, corresponding to the case in Fig. 4, the accuracy of the GEM model at high frequencies is more apparent [Fig. 5(b)]. While the Wu and Lewin models become inaccurate for $a/\lambda>0.14$ and 0.17, respectively, the GEM model maintains accuracy up to $a/\lambda\approx0.34$, doubling the high-frequency limit of the existing EM theories. Moreover, the GEM model has the lowest cross section throughout most of the Mie-resonance region (i.e., for $0.3<a/\lambda<0.5$). Note that the accuracy of the GEM model is limited by numerical error in solving Eq. (2), and that better accuracy is achievable by decreasing the tolerance of the root-finding algorithm.

\section{\label{sec:level1}Summary}
In summary, we have developed a self-consistent, generalized effective-medium model for metamaterials containing spherical particles. The model adequately captures the physics of spatial dispersion and predicts antiresonances and a finite effective refractive index at the resonances. The model provides significant improvement over existing effective-medium theories, particularly close to the Mie resonances and over a broader range of the constituitive permittivities, while retaining their mathematical clarity. These salient features suggest that the GEM model can be a reliable and efficient tool for modeling and designing metamaterials for novel applications.
%\begin{acknowledgments}

This work was funded by ONR (Program Manager, Dr. Mark Spector) through Grant No. N00014-12-1-0722.
%\end{acknowledgments}

\bibliography{bib}% Produces the bibliography via BibTeX.
\end{document}